\documentstyle[a4,epsf,12pt]{article}
\begin{document}
\sloppy

\titlepage
\begin{flushright}
November 2000
\end{flushright}
\vskip 1cm
\begin{center}
{\bf \Large New exact solutions for inflationary cosmological perturbations}
\end{center}

\vskip 1cm

\begin{center}
{\large J\'er\^ome Martin$^{a}$\footnote{email: jmartin@iap.fr} $\&$ 
Dominik J.~Schwarz$^{b}$\footnote{email: dschwarz@hep.itp.tuwien.ac.at}}
\end{center}
\vskip 0.5cm
\begin{center}
$^a$ {\it DARC, Observatoire de Paris-Meudon, \\
UMR 8629 CNRS, 92195 Meudon Cedex, France.}\\
$^b$ {\it Institut f\"ur Theoretische Physik, Technische Universit\"at Wien,\\
Wiedner Haupstra\ss e 8--10, 1040 Wien, Austria.}
\end{center}

\vskip 2cm
\begin{center}
{\large Abstract}
\end{center}
\noindent
{}From a general ansatz for the effective potential of cosmological 
perturbations we find new, exact solutions in single-scalar-field inflation: 
a three parameter family of exact inflationary solutions that encompasses 
all exact solutions that have been known previously (power-law inflation, 
Easther's model, and a generalised version of Starobinsky's solution). The 
main feature of this new family is that the spectral indices are scale 
dependent.

\newpage

\section{Introduction}

The observation of the first acoustic peak in the spectrum of anisotropies 
in the CMBR (Cosmic Microwave Background Radiation) \cite{B98a,MAX1a} leaves
inflation as the only mechanism that can provide seeds for the observed 
large-scale structure of the Universe. The theory of inflation predicts 
primordial density perturbations \cite{scalar} and primordial gravitational 
waves \cite{tensor} from quantum fluctuations of the vacuum during the 
inflationary epoch. So far, inflationary predictions have been based either 
on the slow-roll approximation \cite{SL} or on power-law inflation 
\cite{AW}. However, it is clear that inflation might 
occur in a much broader context and we should discriminate between 
generic predictions of inflation and predictions of a specific
scenario. 
\par
For any model of inflation, the scalar power spectrum 
can be expressed as a Taylor series around some pivot scale 
$k_*$ \cite{KT,CGL}
\begin{equation}
\label{TaylorP}
\ln (k^3P)=\ln (k^3_*P_*)+(n_*-1)\ln \biggl(\frac{k}{k_*}\biggr) +
\frac{1}{2}\frac{{\rm d}n}{{\rm d}\ln k}\biggl \vert _{k_*}
\ln ^2\biggl(\frac{k}{k_*}\biggr)+\cdots .
\end{equation}
A similar expression can be written for the power spectrum of 
tensor perturbations. The quadratic term represents the first 
deviation of the spectrum from a power-law shape and is often 
referred as the ``running'' of the spectral index $n$. No presently 
known exact solutions \cite{AW,E} gives rise to a ``running'' 
spectral index. In this letter, we present a class of exact 
solutions which do exhibit $k$-dependent spectral 
indices. 
\par
It is difficult to predict the coefficients of the 
series (\ref{TaylorP}) in general. One has to rely either 
on approximations or on numerical mode-by-mode integration 
\cite{GL,WMS,CGKL}. The best currently 
available analytical method is the slow-roll approximation 
\cite{SL,Lea} where ``running'' only shows up at second order in 
the slow-roll parameters \cite{KT,CGL}. The precision of any 
approximation scheme should be controlled, this 
means that the errors should quantified. This can only be done by 
comparing the approximate power spectrum to an exact one. For 
inflationary models without ``running'' of the spectral indices, 
we have already analysed this question for the slow-roll 
approximation up to the first order by comparing to the exact 
solutions of power-law inflation \cite{MS2,MRS}. Therefore, to go 
beyond the first order and to test the accuracy of methods that 
estimate the ``running'' of the spectral index, exact 
solutions with $k$-dependent spectral indices 
are needed. So far, only numerical solutions are available to 
perform an error analysis model by model for the second order of 
the slow-roll approximation and for models where the slow-roll 
conditions are not satisfied \cite{GL,WMS,CGKL}. The 
exact solutions presented here constitute a first step towards 
a more complete analysis.
\par
The generic features of inflation might be studied from two sides: from 
the particle physics point of view the starting point is the inflaton 
potential $V(\varphi)$, together with initial conditions for $\varphi$
and $\dot{\varphi}$. From the cosmological side, one might start from the 
effective potential (or effective action) for the evolution of cosmological
perturbations, together with initial conditions for the Hubble rate $H$
and its time derivative. Here we choose the latter approach, which was 
also taken in Refs.~\cite{E,StaroExact}. We find a family of exact inflationary
solutions for the mode equations of cosmological perturbations that includes 
all previously known solutions as limiting cases and that allows us to gain 
new insight into the generic properties of inflation. 
\par
Let us now introduce some basic tools for the study of cosmological 
perturbations. During inflation, density (scalar) perturbations 
and gravitational waves (tensor perturbations) can be characterised by 
the quantities $\mu _{\rm S,T}(\eta )$ \cite{MFB}, which obey the 
equations (we use the notation of \cite{MS,MS2}):
\begin{equation}
\label{eqmotion}
\mu _{\rm S,T}''+
\biggl[k^2-\frac{z_{\rm S,T}''}{z_{\rm S,T}}\biggr]\mu _{\rm S,T}=0,
\end{equation}
where a prime denotes differentiation with respect to conformal 
time $\eta$ and $z_{\rm S,T}$ are functions of the scale factor 
$a(\eta )$ and of its derivatives only:
\begin{equation}
\label{defz}
z_{\rm S}(\eta )\equiv a\sqrt{\gamma }, \quad z_{\rm T}(\eta )\equiv a,
\end{equation}
with $\gamma (\eta )\equiv 1-(a^2/a^{'2})(a'/a)'$, which is nothing 
but the slow-roll parameter $\epsilon \equiv -\dot{H}/H^2$. Inflation 
occurs when $\gamma <1$. Eq.~(\ref{eqmotion}) 
may be viewed as a ``time-independent'' Schr\"odinger equation with an 
effective potential given by
\begin{equation}
\label{epot}
U_{\rm S,T}(\eta )\equiv {z_{\rm S,T}''\over z_{\rm S,T}}.
\end{equation} 
The assumption that 
the quantum fields are initially placed in the vacuum state, when the 
mode $k$ is subhorizon, fixes the initial conditions for the quantities 
$\mu _{\rm S,T}$. They read:
\begin{equation}
\label{ini}
\lim _{k/(aH)\rightarrow +\infty }\mu _{\rm S,T}(\eta )=
\mp 4\sqrt{\pi }l_{\rm Pl} \frac{e^{-ik(\eta -\eta _{\rm i})}}{\sqrt{2k}},
\end{equation}
where $\eta _{\rm i}$ is an arbitrary initial 
time at the beginning of inflation. Then, an integration of 
Eq.~(\ref{eqmotion}) allows the determination of the 
power spectra $P(k)$. For density perturbations we choose to work with 
the power spectrum of the hypersurface independent quantity $\zeta$, which 
corresponds to the perturbation of the intrinsic curvature of a spatial, 
uniform density hypersurface. For gravitational waves we determine the power 
spectrum of the amplitude $h$. These power spectra are calculated 
according to: 
\begin{equation}
\label{defspec}
k^3P_{\zeta }\equiv \frac{k^3}{8\pi ^2}\biggl\vert\frac{\mu_{\rm S}}{z_{\rm S}}
\biggr \vert ^2, \quad k^3 P_h \equiv \frac{2k^3}{\pi ^2}\biggl \vert 
\frac{\mu _{\rm T}}{z_{\rm T}}\biggr \vert ^2.
\end{equation}
Measurements of the CMBR anisotropies and of the large-scale 
structure probe scales which were well beyond the Hubble 
radius at the end of inflation. Thus, we are interested in the 
modes which satisfy $k/(aH)\ll 1$ at the end of inflation.
It is easy to see from Eq. (\ref{eqmotion}) that the spectra are time 
independent in this regime since $\mu_{\rm S,T} \propto z_{\rm S,T}$ 
once the subdominant mode can be neglected. Let us also note that the 
spectrum of the quantity $\zeta $ is related to the spectrum of the 
Bardeen potential today by $k^3P_{\zeta }=(25/9)k^3P_{\rm \Phi}$. In 
agreement with Eq. (\ref{TaylorP}) the spectral 
indices are defined by: 
$n_{\rm S}-1\equiv {\rm d}\ln (k^3P_{\zeta })/{\rm d}\ln k$ and 
$n_{\rm T}\equiv {\rm d}\ln (k^3P_{h})/{\rm d}\ln k$.
\par
The behaviour of cosmological perturbations during inflation is 
completely fixed once the functions $z_{\rm S,T}(\eta)$ are 
specified. The effective potential $U_{\rm S,T}(\eta)$ follows from
Eq.~(\ref{epot}) and Eq.~(\ref{eqmotion}) can be solved. Integration 
of equation (\ref{defz}) provides the scale factor. Instead of assuming 
specific functions $z_{\rm S,T}(\eta)$, one can also start from the 
effective potential $U_{\rm S,T}(\eta)$ itself, such that 
Eq.~(\ref{eqmotion}) may be solved analytically. In this case the functions 
$z_{\rm S,T}(\eta )$ are known explicitly since $z_{\rm S,T}(\eta )
=\mu_{\rm S,T}(k=0,\eta )$. 
\par
Let us now discuss the shape of $U_{\rm S,T}(\eta )$. One expects that the 
effective potential possesses the following properties: 
$\lim _{|\eta |\rightarrow +\infty }U_{\rm S,T}(\eta )\ll k^2$, for any 
wavenumber of interest, in order for the modes to be inside the horizon at 
the initial time $\eta _{\rm i}$ (so that we can fix the normalisation from 
quantum-mechanical considerations); it seems also reasonable 
to assume that $\lim _{|\eta |\rightarrow 0}U_{\rm S,T}(\eta )\gg k^2$ which 
guarantees that the quantum fluctuations are amplified (i.e. are frozen 
outside the horizon). These requirements do not single out a unique effective 
potential. However, a general simple ansatz satisfying these conditions is 
given by:
\begin{equation}
\label{generalshape}
U_{\rm S,T}(\eta )=\sum _{m=1}^{M}\frac{c_m}{|\eta |^m},
\end{equation}
where the $c_m$'s should be determined for each specific model of 
inflation. It is necessary to have $c_{\rm min}$ and $c_M$ 
positive. The coefficient $c_1$ defines 
a characteristic scale $k_{\rm C}\approx c_1$. The corresponding term 
in the series (\ref{generalshape}) dominates the other ones when 
$|\eta| \rightarrow \infty $ and will therefore determine the power 
spectrum at very large scales. 
\par
Inflationary models for which the evolution of cosmological perturbations
has been studied so far, power-law inflation and slow-roll inflation, 
are chosen such that $c_m=0$ if $m\neq 2$. In this sense they are special 
models. In particular, they lead to $k$-independent 
spectral indices. 
\par
In this article we study the behaviour of cosmological 
perturbations for models characterised by an effective 
potential given in Eq.~(\ref{generalshape}). A first step in the analysis 
of these models consists in studying the effective potential
\begin{equation}
\label{shapeexact}
U_{\rm S,T}(\eta )=\frac{c_1}{|\eta |} + \frac{c_2}{|\eta |^2}.
\end{equation}
The final power spectra depend on the free parameters $c_1$ and $c_2$ plus 
one parameter contained in $z_{\rm S,T}(\eta)$. We demonstrate that the 
general solution for this potential is given in terms of Whittaker functions 
and we show explicitly how all the previously known cases can be recovered 
from this more general solution.
\par
The initial conditions together with Eq.~(\ref{shapeexact}) determine 
the region of the inflaton potential explored during the evolution of the 
field. We always take $\gamma_{\rm i} < 1$, such that an 
inflationary phase is guaranteed. Although all the properties of the 
cosmological perturbations can be calculated analytically, it is not possible 
to determine the corresponding inflaton potential analytically. This issue is 
important since we would like to know whether the new class of solutions 
corresponds to generic potentials $V(\varphi)$ from the particle physics 
point of view. Therefore, we calculate the corresponding inflaton potentials 
numerically for the scalar case, with initial conditions inspired by chaotic 
inflation.  
\par
This letter is organised as follows. In section II, we briefly remind 
the reader about known analytical solutions. In section III, we present 
a new family of exact solutions and in section IV we study various 
limits of the parameters of the new solutions.

\section{Known analytical solutions}

In the simplest case all coefficients $c_m$ vanish and $U_{\rm S,T}=0$ 
(Easther's solution \cite{E}). The corresponding functions $z_{\rm S,T}$ 
are given by $z_{\rm S,T}(\eta )\equiv B\eta +A$, where $A$ and $B$ 
are free parameters. For the tensor sector, this gives just the radiation 
dominated Universe. For the scalar sector, the general solution of 
Eq.~(\ref{eqmotion}) can be written as $\mu_{\rm S}(\eta )= 
C_1 e^{ik\eta } +C_2 e^{-ik\eta }$, where $C_1$ and $C_2$ are two arbitrary 
constants to be determined by the initial conditions (\ref{ini}). They read 
$C_1=0$ and $C_2 = - 4\sqrt{\pi}l_{\rm Pl} e^{ik\eta _{\rm i}}/\sqrt{2k}$.
In the large scale limit we have $z_{\rm S} \approx A$ and the 
spectrum of density perturbations reads:
\begin{equation}
\label{Espec}
k^3P_{\zeta }=\frac{l_{\rm Pl}^2}{\pi A^2}k^2.
\end{equation}
The spectral index is given by $n_{\rm S}=3$, which is in obvious 
contradiction with observations. The analytic form of the 
corresponding scalar potential $V(\varphi)$ has been given in Ref.~\cite{E}. 
\par
All other cases known so far assume a potential such that 
$U_{\rm S,T}=\alpha (\alpha +1)/\eta ^2$, i.e. $c_m=0$ if $m\neq 2$ and 
$c_2=\alpha (\alpha +1)$, where $\alpha$ is a free parameter \cite{MS2}. 
The corresponding functions $z_{\rm S,T}$ can be expressed 
as $z_{\rm S,T}(\eta ) = A/|\eta|^\alpha$, where $A$ is another 
free parameter. Thus, we have a two-parameter family characterised 
by $A$ and $\alpha $. For tensor perturbations all these models correspond
to power-law inflation \cite{AW}. For scalar perturbations the subclass 
$\alpha =1$ was studied by Starobinsky \cite{StaroExact}. Another 
subclass is $A=l_0\sqrt{(\alpha -1)/\alpha }$, which corresponds to 
power-law inflation \cite{AW} with the scale factor given by 
$a(\eta )=l_0|\eta |^{-\alpha}$. The quantity $l_0$ has the dimension of a 
length. In this model inflation occurs if $\alpha \ge 1$. In the de Sitter 
case ($\alpha = 1$), $l_0$ is simply the constant Hubble radius during 
inflation. Only for power-law inflation scalar and tensor perturbations can 
be solved analytically at the same time. For scalar perturbations the 
general two-parameter family is not equivalent to power-law inflation.
Of course, $a(\eta )\propto |\eta |^{-\alpha }$ is a particular 
solution of $z_{\rm S}(\eta ) = A/|\eta|^\alpha $, but it is not the 
general solution. The limit $A$ to zero and $\alpha$ close to one 
reproduces a slow-roll inflation model. Eq.~(\ref{eqmotion}) is 
solved in terms of Bessel functions,
$\mu_{\rm S}(\eta )=(k\eta )^{\frac{1}{2}}[C_1J_{\alpha + 
\frac{1}{2}}(k\eta )
+C_2J_{-\alpha  - \frac{1}{2}}(k\eta )]$,
where and $C_1$ and $C_2$ are constants fixed from Eq.~(\ref{ini}). 
The spectrum may be calculated exactly to read
\begin{equation}
\label{zetaexact}
k^3 P_{\zeta}(k) = {l_{\rm Pl}^2\over \pi ^2 A^2}  
2^{2\alpha} \Gamma ^2\biggl(\alpha + \frac{1}{2}\biggr)k^{-2(\alpha -1)} \ .
\end{equation}
The corresponding spectral index is $n_{\rm S}=3-2\alpha $. In particular, 
we have $k^3P_{\zeta }=l_{\rm Pl}^2/(\pi A^2)$ and $n_{\rm S}=1$ 
for $\alpha =1$. For gravitational waves the same ansatz gives a power-law 
model with spectral index $n_{\rm T}=2-2\alpha$. 

\begin{figure}
\setlength{\unitlength}{\linewidth}
\begin{picture}(1,0.45)
\put(0.16,0.28){\makebox(0,0){$V$}}
\put(0.5,0.03){\makebox(0,0){$\varphi/\varphi_{\rm i}$}}
\put(0.5,0.25){\makebox(0,0){\epsfig{figure=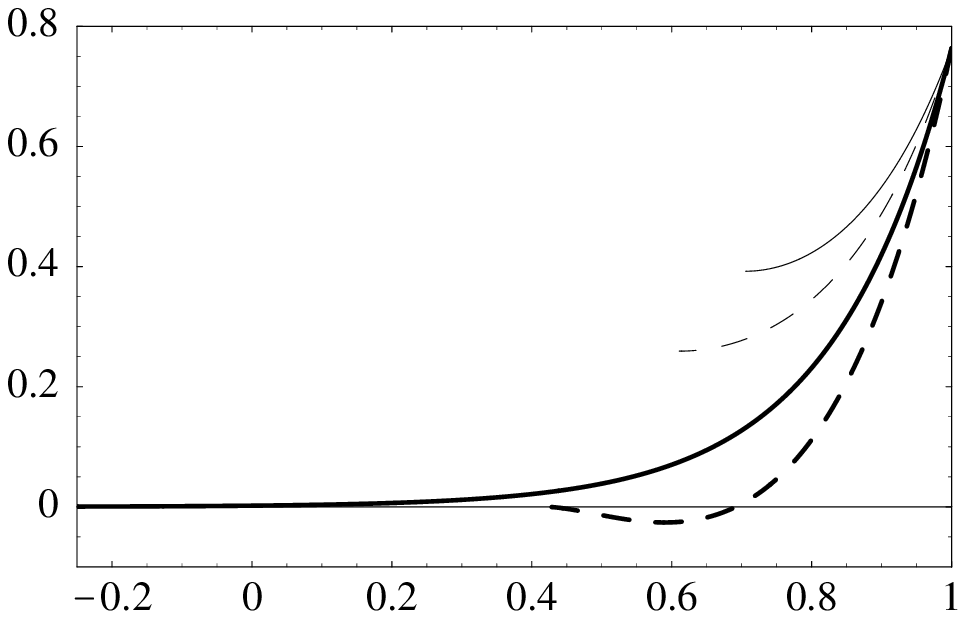,width=0.6\linewidth}}}
\end{picture}
\caption{The generalised Starobinsky potentials for $\alpha = 
1,2,\alpha_{\rm pl}\equiv -\beta -1,5$ (from top to bottom) and $A = 1$, where 
$\alpha_{\rm pl} \approx 3.44$. The initial conditions
are $\gamma_{\rm i} = (\alpha_{\rm pl} - 1)/\alpha_{\rm pl} \approx 0.7$ 
and $H_{\rm i} = \sqrt{8\pi G/3}$.}
\label{potPL}
\end{figure}
The inflaton potential for the scalar sector is displayed in Fig.~(\ref{potPL})
for various values of $\alpha $. We use initial conditions motivated by  
the scenario of chaotic inflation, i.e.~the potential and the kinetic energy 
densities of the inflaton field are of the order of the Planck energy density
initially. The condition $\gamma _{\rm i}<1$ guarantees that inflation 
takes place. For power-law inflation, the potential is given by: 
$V(\varphi )=V_{\rm i}\exp [4\sqrt{\pi \gamma }(\varphi 
-\varphi _{\rm i})/m_{\rm Pl}]$. 

\section{A new analytical solution}

We now turn to the main result of this article. The function 
$z_{\rm S,T}$ corresponding to the effective potential (\ref{shapeexact}) 
can be written as :
\begin{equation}
\label{Wz}
z_{\rm S,T}(\eta) = {2A c_1^{\xi -1/2}\over \Gamma(2\xi)}
\sqrt{c_1\vert\eta\vert} K_{2\xi}(2\sqrt{c_1\vert\eta\vert})\ ,
\end{equation}
where $A>0$ and $\xi \equiv \sqrt{c_2+1/4}\ge 1/2$. $K_{2\xi}$ is a modified
Bessel function of order $2\xi$. We 
now define $\tau $ as $\tau \equiv 2ik\eta $ and $\kappa $ as 
$\kappa \equiv -ic_1/(2k)$. $c_1$ defines 
a characteristic scale given by $k_{\rm C}=c_1/2$. Then, the 
equation of motion (\ref{eqmotion}) can be expressed as:
\begin{equation}
\label{Weqmot}
\frac{{\rm d}^2\mu _{\rm S,T}}{{\rm d}\tau ^2}+
\biggl[-\frac{1}{4}+\frac{\kappa }{\tau }
+\frac{(1/4-\xi ^2)}{\tau ^2}\biggr]\mu _{\rm S,T}=0.
\end{equation}
The general solution to this equation is given in terms of Whittaker 
functions 
\begin{equation}
\label{Wsol}
\mu _{\rm S,T}(\tau )=C_1 W_{\kappa,\xi }(\tau )+C_2 W_{-\kappa,\xi}(-\tau).
\end{equation}
The constants $C_1$ and $C_2$ are fixed by the initial conditions. Using 
the asymptotic behaviour of the Whittaker functions in the small-scale 
limit, $\lim _{\vert \tau \vert \rightarrow \infty}
W_{\kappa ,\xi }(\tau )=e^{-\frac{\tau }{2}} \tau^\kappa $, Eq. (9.227) 
of Ref.~\cite{GR}, we find:
\begin{equation}
\label{Wini}
C_1=\mp 4\sqrt{\pi}l_{\rm Pl}\frac{e^{ik\eta_{\rm i}+\pi c_1/(4k)}}{\sqrt{2k}}, 
\quad C_2=0\ .
\end{equation}
Expressing the Whittaker functions in terms of Kummer functions, 
Eqs.~(9.220) of Ref.~\cite{GR}, we deduce that, if $-\xi +1/2 < 0$, 
the growing mode in the large scale limit is given by: 
$\lim _{|\tau |\rightarrow 0}W_{\kappa ,\xi }(\tau )=
\Gamma(2\xi)/[\Gamma (1/2+\xi -\kappa )] (\tau )^{-\xi +1/2}e^{-\tau /2}$. 
Using that $z_{\rm S,T}(\eta) \approx A|\eta|^{\frac{1}{2}-\xi}$ in 
this limit, it is straightforward to calculate the spectrum of density 
perturbations:
\begin{equation}
\label{Wspec}
k^3P_{\zeta}=\frac{l_{\rm Pl}^2}{\pi A^2}\frac{\Gamma ^2(2\xi )}{2^{2\xi -1}}
\frac{k^{3-2\xi }e^{\pi k_{\rm C}/k}}
{\vert \Gamma (1/2+\xi + i k_{\rm C}/k)\vert ^2}.
\end{equation}
This spectrum corresponds to a new exact solution which contains all the 
previously known cases as particular cases, see the next 
section. Using Eq. (8.328.1) of Ref.~\cite{GR}, we see that for  
$k\ll k_{\rm C}$ one has 
$\vert\Gamma (1/2+\xi + i k_{\rm C}/k)\vert^2 \approx 
2\pi \exp(-\pi k_{\rm C}/k)(k_{\rm C}/k)^{2\xi}$. 
This implies that in the 
large-scale limit the spectrum is given by:
\begin{equation}
\label{Wspeclimit1}
k^3P_{\zeta }\approx \frac{l_{\rm Pl}^2k^3}{\pi ^2 A^2}\frac{\Gamma ^2(2\xi )}
{(2 k_{\rm C})^{2\xi }}e^{2 \pi k_{\rm C}/k},
\end{equation}
if $c_1 \neq 0$ and $c_0 = 0$. We see that it is not analytic in the 
region of small $k$. Of course, such an infra-red divergence of the spectrum
is excluded from observation. However, the validity of our description fails 
for modes that leave the horizon at the Planck epoch, i.e.~at the beginning of
inflation in the scenario of chaotic inflation. This sets a natural 
infra-red cut-off to the power spectrum (\ref{Wspec}). In the small-scale 
limit $k\gg k_{\rm C}$ the spectrum tends to 
\begin{equation}
\label{Wspeclimit2}
k^3P_{\zeta }\approx {l_{\rm Pl}^2\over \pi ^2 A^2}  
2^{2\xi -1} \Gamma ^2(\xi )k^{3-2\xi } \ .
\end{equation}
This is the spectrum of the two parameter family studied in 
the previous section, see Eq. (\ref{zetaexact}). The corresponding 
constant spectral index is given by 
$n_{\rm S}=4-2\xi =4-2\sqrt{c_2+1/4}$ and is 
entirely determined by the coefficient $c_2$. 

\section{Limiting cases}

\subsection{$c_1\neq 0$, $c_2=0$: Coulomb solution}

\begin{figure}
\setlength{\unitlength}{\linewidth}
\begin{picture}(1,0.5)
\put(0.2,0.31){\makebox(0,0){$k^3 P_\zeta(k)$}}
\put(0.53,0.02){\makebox(0,0){$k$}}
\put(0.5,0.28){\makebox(0,0){\epsfig{figure=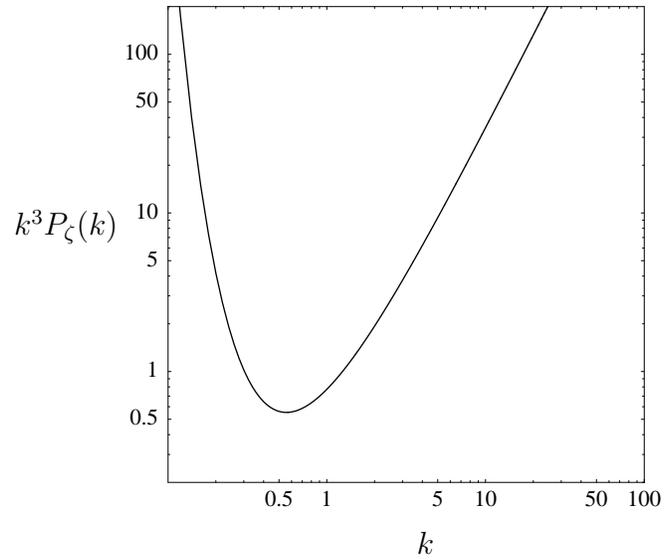,width=0.5\linewidth}}}
\end{picture}
\caption{The scalar power spectrum for the Coulomb case with $A=1$ and 
$c_1= 0.5$.} 
\label{PC}
\end{figure}

\begin{figure}
\setlength{\unitlength}{\linewidth}
\begin{picture}(1,0.45)
\put(0.16,0.28){\makebox(0,0){$V$}}
\put(0.5,0.03){\makebox(0,0){$\varphi/\varphi_{\rm i}$}}
\put(0.5,0.25){\makebox(0,0){\epsfig{figure=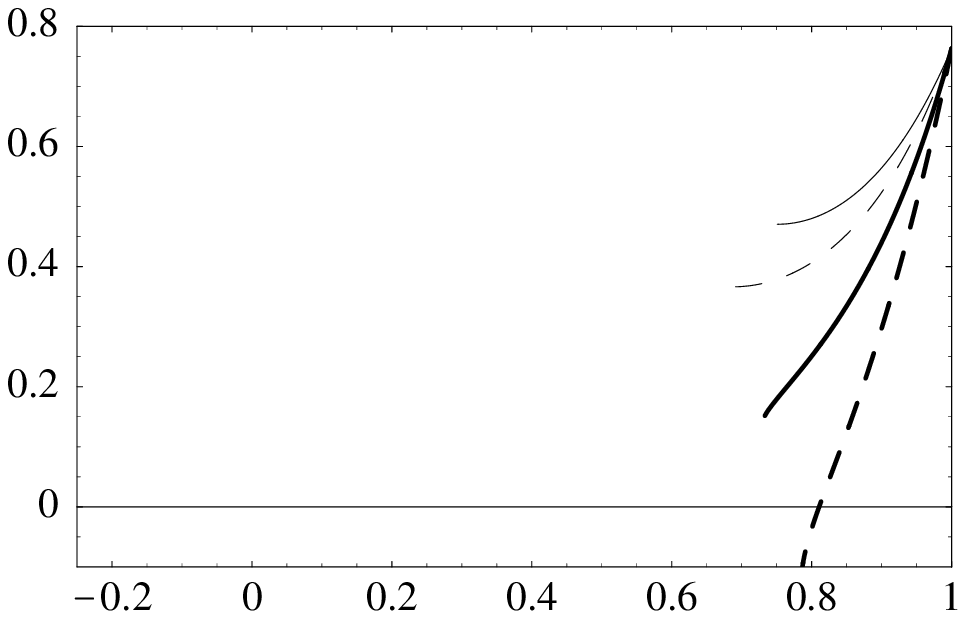,width=0.6\linewidth}}}
\end{picture}
\caption{The potentials for the Coulomb case for $c_1= 0.1, 0.5, 1, 1.5$ and 
$A=1$ with the same initial conditions as before.}
\label{potC}
\end{figure}

As already mentioned, the case usually treated in the literature 
is $c_1=0$, $c_2\neq 0$. On the other hand, the case $c_1\neq 0$, 
$c_2=0$ has never been studied before. Although it is of course just 
a particular case of the general solution given in the previous 
section, it is worth investigating its properties in some details. 
For this purpose, we restart from the beginning and consider the 
following function $z_{\rm S,T}$ 
\begin{equation}
\label{Czsol}
z_{\rm S,T}(\eta )\equiv 2A\sqrt{c_1\vert \eta \vert }
K_1\biggl(2\sqrt{c_1\vert\eta\vert}\biggr).
\end{equation}
The link with Eq. (\ref{Wz}) is obvious since $c_2=0$ corresponds 
to $\xi =1/2$. Let us now define $\rho  $ and $\delta  $ 
according to $\rho \equiv i\tau/2$, $\delta \equiv i\kappa$. Then 
the equation of motion (\ref{eqmotion}) for $\mu _{\rm S,T}(\eta )$ can 
be written as:
\begin{equation}
\label{Ceq}
\frac{{\rm d}^2\mu _{\rm S,T}}{{\rm d}\rho ^2}+
\biggl(1-\frac{2\delta }{\rho }\biggr)\mu _{\rm S,T}=0.
\end{equation}
This equation can be solved in terms of Coulomb functions \cite{AS} with $l=0$, 
$\mu _{\rm S,T}(\rho )=C_1F_0(\delta ;\rho )+C_2G_0(\delta  ;\rho )$.
The coefficients $C_1$ and $C_2$ are fixed by the initial conditions, see 
Eq. (\ref{ini}). We find that $
C_1= \pm 4i\sqrt{\pi }l_{\rm Pl}e^{ik\eta_{\rm i}}/\sqrt{2k},\ 
C_2=-iC_1$.
In the large scale limit, the growing mode is given by the 
irregular Coulomb wave function $G_0(\delta ;\rho)\approx 1/C_0(\delta  )$, 
where $C_0(\delta )\equiv \sqrt{2\pi \delta /(e^{2\pi \delta }-1)}$. Using 
the fact that $K_1(x)\approx 1/x$ when $x$ goes to 
zero, the spectrum is easily derived. For density perturbations, we 
find:
\begin{equation}
\label{Cspeclarge}
k^3P_{\zeta}=\frac{l_{\rm Pl}^2k^3}{2\pi^2 A^2k_{\rm C}}
\biggl(e^{2\pi k_{\rm C}/k}-1\biggr).
\end{equation}
This spectrum is of course nothing but Eq. (\ref{Wspec}) for 
$\xi =1/2$. It is displayed in Fig. (\ref{PC}) for $A=1$ and 
$c_1=0.5$.
In the 
regime where $k\ll k_{\rm C}$ one recovers Eq. (\ref{Wspeclimit1}) whereas if 
$k\gg k_{\rm C}$ this spectrum tends to the Easther's solution 
$k^3P_{\zeta}=(l_{\rm Pl}^2k^2)/(\pi A^2)$ for which $n_{\rm S}=3$ in 
accordance with Eq. (\ref{Wspeclimit2}). The corresponding inflaton 
potentials for various values of $c_1$ are displayed in 
Fig. (\ref{potC}). The initial conditions are chosen as in Fig.~1 
and guarantee that inflation occurs.  

\subsection{$c_1=0$, $c_2\neq 0$: generalised power-law solution}

Putting $c_1=0$ and $\kappa =0$ in Eq. (\ref{Wspec}), one easily 
checks that the result is the generalised power-law 
spectrum, see Eq. (\ref{zetaexact}), with 
$\alpha = \xi -1/2$. The corresponding spectral index is 
$n_{\rm S}=4-2\xi $.

\subsection{$c_1\neq 0$, $c_2=2$}
\begin{figure}
\setlength{\unitlength}{\linewidth}
\begin{picture}(1,0.5)
\put(0.2,0.31){\makebox(0,0){$k^3 P_\zeta(k)$}}
\put(0.53,0.03){\makebox(0,0){$k$}}
\put(0.5,0.28){\makebox(0,0){\epsfig{figure=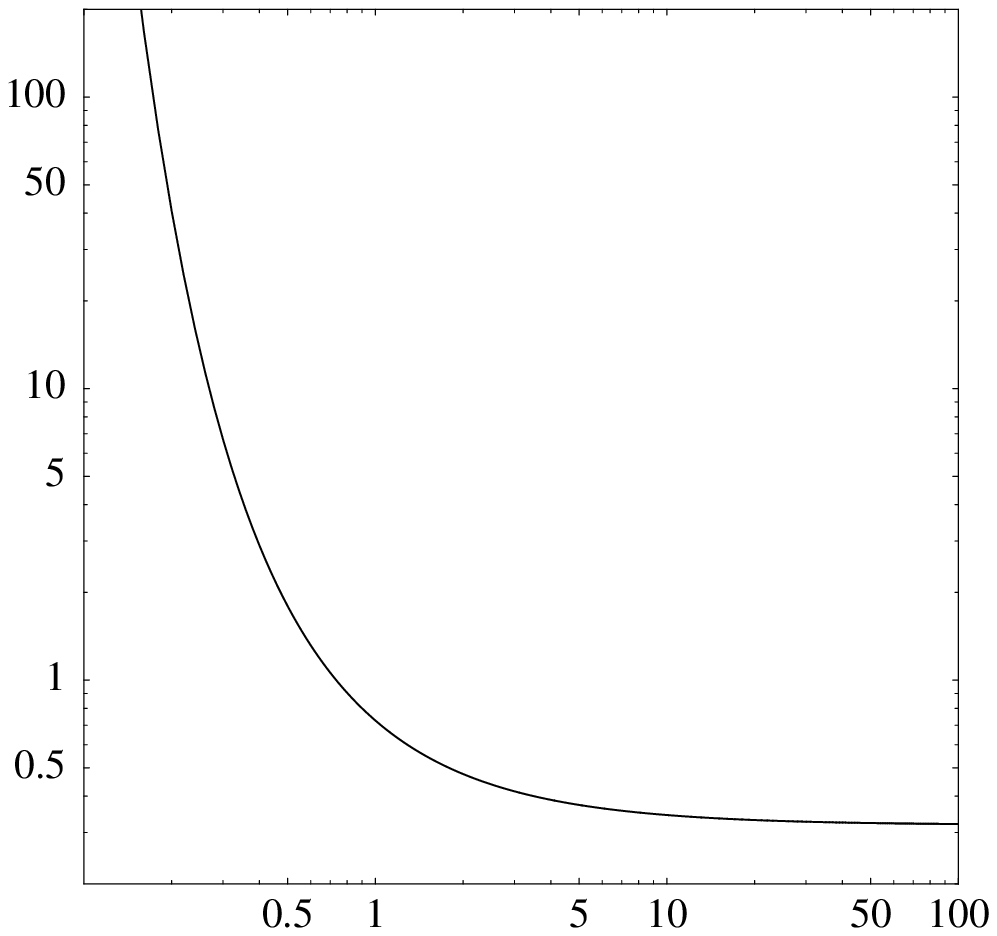,width=0.5\linewidth}}}
\end{picture}
\caption{The scalar power spectrum for the Whittaker case with $A=1, c_2=2$ 
and $c_1 = 0.5$.} 
\label{PW}  
\end{figure}

Let us finally investigate in more details the case $c_2=2$ 
which corresponds to $\xi =3/2$. Using Eq. (8.332.1) of 
Ref.~\cite{GR}, we find that the exact spectrum for density 
perturbations is given by:
\begin{equation}
\label{WspecB=2}
k^3P_{\zeta}^{(c_2=2)}=\frac{l_{\rm Pl}^2}{2\pi ^2A^2}
\frac{k}{k_{\rm C}}\biggl(1+\frac{k_{\rm C}^2}{k^2}\biggr)^{-1}
 \biggl(e^{2\pi k_{\rm C}/k}-1\biggr).
\end{equation}
The spectrum is displayed in Fig. (\ref{PW}) for $A=1$ and $c_1=0.5$. 

\begin{figure}
\setlength{\unitlength}{\linewidth}
\begin{picture}(1,0.45)
\put(0.16,0.28){\makebox(0,0){$V$}}
\put(0.5,0.03){\makebox(0,0){$\varphi/\varphi_{\rm i}$}}
\put(0.5,0.25){\makebox(0,0){\epsfig{figure=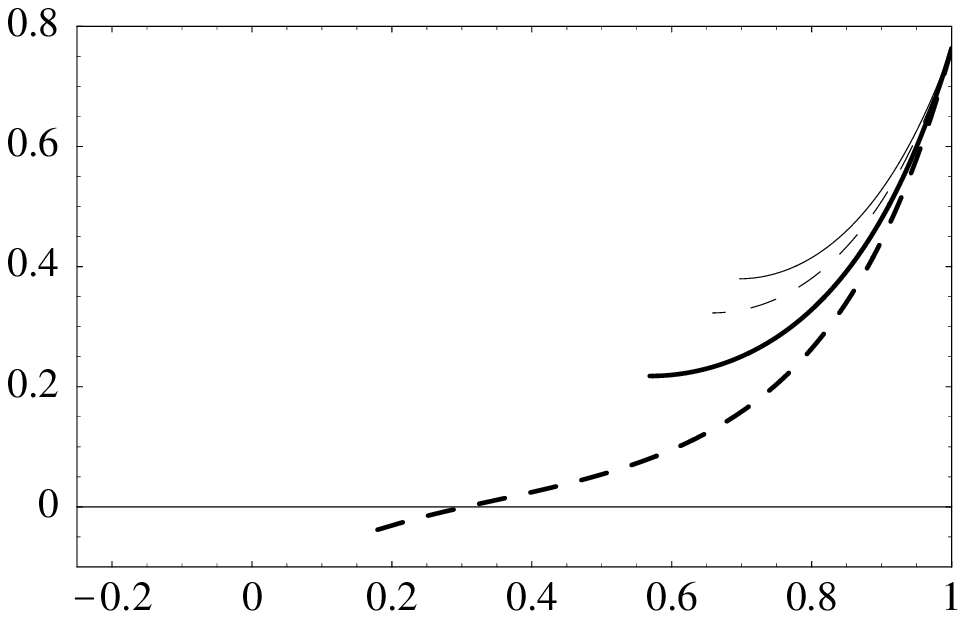,width=0.6\linewidth}}}
\end{picture}
\caption{The potentials for the Whittaker case for $A=1, c_2=2$ and 
$c_1= 0.1, 0.5, 1, 1.5$ with the same initial conditions as before.}
\label{potW}
\end{figure}

In the limit $k_{\rm C}/k\ll 1$, the previous spectrum 
reduces to $k^3P_{\zeta}^{(c_2=2)}=(l_{\rm Pl}^2)/(\pi A^2)$, 
i.e. the spectrum of Starobinsky's solution ($n_{\rm S}=1$)
as expected. The potential of the inflaton is displayed 
in Fig.~(\ref{potW}) for various values of $c_1$ with the 
same initial conditions as in Figs.~1 and 3. 

We briefly conclude in recalling that the main result of this article is 
the discovery of a new family of exact solutions for inflationary cosmological 
perturbations which encompasses all the previously known cases as 
limiting cases, see Table I. This new family could help to shed light 
on inflationary models which do not fulfill the slow-roll 
conditions. Moreover, it can be used to test methods of approximation 
that are valid for models possessing  ``running'' of the spectral 
indices. For this purpose the knowledge of exact solutions allows us to avoid 
a model-by-model analysis, as is necessary for numerical methods. In this 
spirit, the precision of the slow-roll approximation was recently investigated 
using power-law inflation as an exact solution \cite{MS2,MRS}.

\begin{table}
\begin{tabular}{c c c c}
\hline \\
$c_1$ & $c_2$ & Regime  & $k^3P_{\zeta }$ \\ \hline \hline 
\\
$\neq 0$ & $\neq 0$ & Exact & Whittaker's solution (\ref{Wspec}), 
$n_{\rm S}=n_{\rm S}(k)$ \\

$\neq 0$ & $\neq 0$ & $k\ll k_{\rm C}$ & Eq. (\ref{Wspeclimit1}), 
$n_{\rm S}=n_{\rm S}(k)$ \\

$\neq 0$ & $\neq 0$ & $k\gg k_{\rm C}$ & Generalised power-law (\ref{zetaexact}), 
$n_{\rm S}=4-2\sqrt{c_2+1/2} $ \\

$0$ & $0$ & Exact & Easther's solution (\ref{Espec}), $n_{\rm S}=3$ \\

$0$ & $\neq 0$ & Exact & Generalised power-law (\ref{zetaexact}), 
$n_{\rm S}=4-2\sqrt{c_2+1/2} $ \\

$\neq 0$ & $0$ & Exact & Coulomb's solution (\ref{Cspeclarge}), 
$n_{\rm S}=n_{\rm S}(k)$ \\

$\neq 0$ & $0$ & $k\gg k_{\rm C}$ & Easther's solution (\ref{Espec}), $n_{\rm S}=3$ \\

$\neq 0$ & $2$ & Exact & Eq. (\ref{WspecB=2}), $n_{\rm S}=n_{\rm S}(k)$ \\

$\neq 0$ & $2$ & $k\gg k_{\rm C}$ & Starobinsky's solution (\ref{zetaexact}), 
$n_{\rm S}=1$ \\
\hline 

\label{summary}
\end{tabular}
\caption{Spectra for different values of $c_1$ and $c_2$ and and in 
different regimes.}
\end{table}

\section*{Acknowledgement}

D.J.S.~would like to thank the Austrian Academy of Sciences for financial 
support. J.M.~would like to thank the ITP, TU Wien (Vienna, Austria) for 
warm hospitality.

\end{document}